\documentclass[reqno,11pt]{article}
\usepackage{psfig, amsmath, amstexnb, amsthm}
\usepackage{amssymb}  

\input{stochbif.sty}
\def\captionspace{\vspace{2mm}}	

\begin{document}


\title{Metastability in simple climate models:\\ 
       Pathwise analysis of slowly driven Langevin equations} 

\author{Nils Berglund and Barbara Gentz}
\date{}   

\maketitle

\begin{abstract}
\noindent
We consider simple stochastic climate models, described by slowly
time-dependent Langevin equations. We show that when the noise
intensity is not too large, these systems can spend substantial
amounts of time in metastable equilibrium, instead of adiabatically
following the stationary distribution of the frozen system. This
behaviour can be characterized by describing the location of typical
paths, and bounding the probability of atypical paths. We illustrate
this approach by giving a quantitative description of phenomena
associated with bistability, for three famous examples of
simple climate models: Stochastic resonance in an energy balance model
describing Ice Ages; hysteresis in a box model for the Atlantic
thermohaline circulation; and bifurcation delay in the case of the
Lorenz model for Rayleigh--B\'enard convection. 
\end{abstract}

\leftline{\small{\it Date.\/} November 4, 2001.}
\leftline{\small 2000 {\it Mathematical Subject Classification.\/} 
37H20 (primary), 60H10, 34E15, 82C31 (secondary).}
\noindent{\small{\it Keywords and phrases.\/}
Stochastic resonance, dynamical hysteresis, bifurcation delay,
double-well potential, first-exit time, scaling laws, Lorenz
model, thermohaline circulation, white noise, coloured noise.}  


\section{Introduction}
\label{sec_in}

One of the main difficulties of realistic climate models is that they
involve a huge number of interacting degrees of freedom, on a wide range
of  time and length scales. In order to be able to control these models
analytically, or at least numerically, it is necessary to simplify them by
eliminating the less relevant degrees of freedom (e.\,g.\ high-frequency or
short-wavelength modes). A possible way to do this is to average the
equations of motion over all fast degrees of freedom, a rather drastic
approximation. As proposed by Hasselmann~\cite{Hasselmann} (see also
\cite{Arnold2}), a more realistic approximation is obtained by modeling
the effect of fast degrees of freedom by noise. 

In a number of cases, it is appropriate to distinguish between three rather
than two time scales: Fast degrees of freedom (e.\,g.\ the \lq\lq
weather\rq\rq), which are modeled by a stochastic process; intermediate
\lq\lq dominant modes\rq\rq\ (e.\,g.\ the average temperature of the
atmosphere) whose dynamics we want to predict; and slow degrees of freedom
(e.\,g. the mean insolation depending on the eccentricity of the Earth's
orbit), which evolve on very long time scales of several centuries or
millennia, and can be viewed as an external forcing. Such a system can
often be modeled by a slowly time-dependent Langevin equation
\begin{equation}
\label{in1}
\6x_t = f(x_t,\eps t)\6t + \sigma G(\eps t)\6W_t,
\end{equation}
where the adiabatic parameter $\eps$ and the noise intensity $\sigma$ are
small parameters, $W_t$ is a standard vector-valued Wiener process
(describing white noise) and $G$ is a matrix. 

Our aim in this paper is to describe the effect of the noise term on the
dynamics of~\eqref{in1}, assuming the dynamics without noise is known. For
this purpose, we will concentrate on bistable systems, which frequently
occur in simple climate models: For instance, in models for the major Ice
Ages, where the two possible stable equilibria correspond to warm and cold
climate~\cite{BPSV}, or in models of the Atlantic thermohaline circulation
\cite{Stommel,Rahmstorf}. Noise may enable transitions between the two
stable states, which would be impossible in the deterministic case, and our
main concern will be to quantify this effect.  

The method used to study the stochastic differential equation
(SDE)~\eqref{in1} will depend on the time scale we are interested in. Let us
first illustrate this on a static one-dimensional example, namely the
overdamped motion in a symmetric double-well potential:
\begin{equation}
\label{in2}
\6x_t = -\dpar{}x V(x_t) \6t + \sigma\6W_t,
\qquad
V(x) = \frac14 b x^4 - \frac12 a x^2.
\end{equation}
where $a$ and $b$ are positive constants. The potential has two wells at
$\pm\sqrt{a/b}$, separated by a barrier of height $H=a^2/(4b)$. A first
possibility to analyse this equation is to compute the probability density
$p(x,t)$ of $x_t$. It obeys the Fokker--Planck equation
\begin{equation}
\label{in3}
\dpar{}{t} p(x,t) = \dpar{}{x} \biggbrak{\dpar Vx(x) p(x,t)} + 
\frac{\sigma^2}2 \dpar{^2}{x^2} p(x,t),
\end{equation}
which admits in particular the stationary solution 
\begin{equation}
\label{in4}
p_0(x) = \frac1N \e^{-2V(x)/\sigma^2},
\end{equation}
where $N$ is the normalization. At equilibrium, there is equal probability
to find $x_t$ in either potential well, and for weak noise it is unlikely
to observe $x_t$ anywhere else than in a \nbh\ of order $\sigma$ of one of
the wells. 

Assume now that the initial distribution $x_0$ is concentrated at the
bottom $\sqrt{a/b}$ of the right-hand potential well. Then it may take
quite a long time for the density to approach its asymptotic value
\eqref{in4}. A possible way to investigate this problem relies on spectral
theory. Denote the right-hand side of~\eqref{in3} as $\cL p(x,t)$, where
$\cL$ is a linear differential operator. The stationary density~\eqref{in4}
is an eigenfunction of $\cL$ with eigenvalue $0$. We may assume that $\cL$ has
eigenvalues $\dots < \lambda_k < \dots < \lambda_2 < \lambda_1 < 0$,
c.\,f.~\cite[Section~6.7]{HorstLefever}. Decomposing $p(x,t)$ on a
basis of eigenfunctions of $\cL$, we see that $p$ approaches the
stationary solution in a characteristic time of order $1/\abs{\lambda_1}$. 

There exists, however, a much more precise description of the process $x_t$
than by its probability density. Recall that for almost every realization
$W_t(\w)$ of the Brownian motion, the sample path $t\mapsto x_t(\w)$ 
is continuous. Instead of computing the time needed for $p(x,t)$ to relax
to $p_0(x)$, we can consider the random variable
\begin{equation}
\label{in5}
\tau(\w) = \inf\bigsetsuch{t>0}{x_t(\w)<0},
\end{equation}
describing the first time at which the path $x_t$ crosses the saddle (one
could as well consider the first time the bottom of the left-hand well is
reached). The distribution of $\tau$ is asymptotically exponential, with
expectation behaving in the weak-noise limit like Kramers' time 
\begin{equation}
\label{in6}
T_{\math{Kramers}} = \e^{2H/\sigma^2}.
\end{equation}
A mathematical theory allowing to estimate first-exit times for general
$n$-dimensional systems (with a drift term not necessarily
deriving from a potential) has been developed by Freidlin and
Wentzell~\cite{FW}. In specific situations, more precise results are
available, for instance subexponential corrections to the asymptotic
expression~\eqref{in6}, see~\cite{Azencott,FJ}. Even the limiting behaviour
of the distribution of the first-exit time from a neighbourhood of a
unique stable equilibrium point has been obtained~\cite{Day1}. The
first-exit time from a neighbourhood of  a saddle has been considered
by Kifer in the seminal paper~\cite{Kifer}.  

If the noise intensity $\sigma$ is small (compared to the square root of
the barrier height), then the time needed to overcome the potential barrier
is extremely long, and the time required to relax to the stationary
distribution $p_0(x)$ is even longer. In fact, on time scales shorter than
Kramers' time, solutions of~\eqref{in2} starting in one potential well will
hardly feel the second potential well. As we will see in
Section~\ref{sec_st}, $x_t$ is well approximated by an Ornstein--Uhlenbeck
process, describing the overdamped motion of a particle in a potential of
constant curvature $c=2a$. The Ornstein--Uhlenbeck process relaxes to a
{\em stationary\/} Gaussian process with variance $\sigma^2/(2c)$ in a
characteristic time 
\begin{equation}
\label{in7}
T_{\math{relax}} = \frac1c.
\end{equation}
Thus for $0\leqs t\ll T_{\math{relax}}$, the behaviour of $x_t$ is
transient; for $T_{\math{relax}} \ll t \ll T_{\math{Kramers}}$, $x_t$ is
close to a stationary Ornstein--Uhlenbeck process with variance
$\sigma^2/(2c)$; and only for $t\gg T_{\math{Kramers}}$ will the
distribution of $x_t$ approach the bimodal stationary solution~\eqref{in4}.
This phenomenon, where a process seems stationary for a long time before
ultimately relaxing to a new (possibly stationary) state, is known as
\defwd{metastability}. It is all the more remarkable in an asymmetric
double-well potential: then a process starting at the bottom of the
shallow well will first relax to a metastable distribution
concentrated in the shallow well, which is radically different from
the stationary distribution having most of its mass concentrated in the
deeper well.  

A different approach, based on the concept of random attractors
(see~\cite{CF1,Schmal,Arnold1}), gives complementary information on the
long-time regime. In particular, in~\cite{CF2} it is proved that for 
arbitrarily weak noise, paths of~\eqref{in2} with different initial
conditions but same realization of noise almost surely converge to a
random point. The time needed for this convergence, however, diverges 
rapidly in the limit $\sigma\to0$, because paths starting in different
potential wells are unlikely to overcome the potential barrier and start
approaching each other before Kramers' time. 

We now turn to situations in which the potential varies slowly in time. For
simplicity, we will consider the family of Ginzburg--Landau potentials 
\begin{equation}
\label{in8}
V(x,\lambda,\mu) = \frac14 x^4 - \frac12 \mu x^2 - \lambda x,
\end{equation}
and let either $\lambda$ or $\mu$ vary in time, with low speed $\eps$. 
For instance, $\lambda$ or $\mu$ may depend periodically on time, with low
frequency $2\pi\eps$. The potential $V$ has two wells if $27\lambda^2
< 4\mu^3$ and one well if $27\lambda^2 > 4\mu^3$, and when 
$\lambda$ or $\mu$ are varied, the number of wells may
change. Crossing one of the curves $27\lambda^2 = 4\mu^3$, $\mu>0$,
corresponds to a saddle--node bifurcation, and crossing the point
$\lambda=\mu=0$ corresponds to a pitchfork bifurcation.  

The slow time-dependence introduces a new time scale
$T_{\math{forcing}}=1/\eps$. Since curvature and barrier height are no
longer constant, we replace the definitions~\eqref{in6} and~\eqref{in7} by 
\begin{equation}
\label{in9}
T_{\math{Kramers}}^{(\math{max})} = \e^{2H_{\math{max}}/\sigma^2}
\qquad\qquad
\text{and}
\qquad\qquad
T_{\math{relax}}^{(\math{min})} = \frac1{c_{\math{max}}},
\end{equation}
where $H_{\math{max}}$ denotes the maximal barrier height during one
period, and $c_{\math{max}}$ denotes the maximal curvature at the bottom of
a potential well. Here we are interested in the regime 
\begin{equation}
\label{in10}
T_{\math{relax}}^{(\math{min})} \ll T_{\math{forcing}} \ll
T_{\math{Kramers}}^{(\math{max})},
\end{equation}
which means that the process has time to reach a metastable \lq\lq
equilibrium\rq\rq\ during one period, but not the bimodal stationary
distribution. Mathematically, we thus assume that $\eps\ll c_{\math{max}}$
and $\sigma^2\ll 2H_{\math{max}}/\abs{\log\eps}$. We allow, however, the
minimal curvature and barrier height to become small, or even to vanish. 

For time-dependent potentials, the Fokker--Planck equation~\eqref{in3} is
even harder to solve (and in fact, it does not admit a stationary
solution). Moreover, random attractors are not straightforward to define in
this time-dependent setting. 
We believe that the dynamics on time scales shorter than
$\smash{T_{\math{Kramers}}^{(\math{max})}}$ is discussed best via an
understanding of \lq\lq typical\rq\rq\ paths. The idea is to show that
the vast majority of paths remain concentrated in small space--time
sets, whose shape and size depend on the potential and the noise
intensity. These sets are typically located in a \nbh\ of the 
potential wells, but under some conditions paths may also switch potential
wells. 
There are thus two problems to solve: first characterize the sets in which
typical paths live, and then estimate the probability of atypical paths. It
turns out that these properties have universal characteristics, depending
only on qualitative properties of the potential, especially its bifurcation
points. 

We start, in Section~\ref{sec_st}, by discussing the simplest situation, 
which occurs when the initial condition of the process lies in the basin of
attraction of a stable equilibrium branch. For sufficiently small noise
intensity, the majority of paths remain concentrated for a long time in a
\nbh\ of the equilibrium branch. We determine the shape of this \nbh\ 
and outline how coloured noise can decrease the spreading of paths.  

Section~\ref{sec_sr} is devoted to the phenomenon of stochastic resonance.
We first recall the energy-budget model introduced in~\cite{BPSV} to give a
possible explanation for the close-to-periodic appearance of the major Ice
Ages. This model is equivalent to the overdamped motion of a particle in a
modulated double-well potential, where the driving amplitude is too small
to allow for transitions between wells in the absence of noise. Turning to
the description of typical paths, we find a threshold value for the noise
intensity below which the paths remain in one well, while above threshold,
they switch back and forth between wells twice per period. The switching
events occur close to the instants of minimal barrier height. 
Several important quantities have a power-law dependence on the small
parameters, in particular the critical noise intensity, the width of
transition windows, and the exponent controlling the exponential decay of
the probability of atypical paths. 

In Section~\ref{sec_h}, we start by discussing a variant~\cite{Cessi} of
Stommel's box model~\cite{Stommel} of the Atlantic thermohaline
circulation. Assuming slow changes in the typical weather, this model  also
reduces to the motion in a modulated double-well potential, where the
modulation depends on the freshwater flux. If the amplitude of the
modulation exceeds a threshold, the potential barrier vanishes twice per
period, so that the deterministic motion displays hysteresis. Additive
noise influences the shape of hysteresis cycles, and may even create
macroscopic cycles for subthreshold modulation amplitude. We characterize
the distribution of the random freshwater flux causing the system to switch
from one stable state to the other one. 

Finally, in Section~\ref{sec_d}, we consider the Lorenz model for
Rayleigh--B\'enard convection with slowly increasing heating. In the
deterministic case, convection rolls appear only some time after the steady
state looses stability in a pitchfork bifurcation. This bifurcation delay
is significantly decreased by additive noise, as soon as its intensity is
not exponentially small. 

\goodbreak


\section{Near stable equilibria}
\label{sec_st}

Let us start by investigating Equation~\eqref{in1} in the one-dimensional
case, i.\,e., when $x_t$, $W_t$ and $G(t)=g(t)$ are scalar. Since we are
interested in the dynamics on the time scale $T_{\math{forcing}}=1/\eps$,
we rescale time by a factor $\eps$, which results in the SDE
\begin{equation}
\label{st1}
\6x_t = \frac1\eps f(x_t,t)\6t + \frac\sigma{\sqrt\eps} g(t)\6W_t. 
\end{equation}
The factor $1/\sqrt\eps$ is due to the diffusive nature of the Brownian
motion. 

In this section, we will consider the dynamics near a stable equilibrium
branch of $f$, i.\,e., a curve $x^\star(t)$ such that 
\begin{equation}
\label{st2}
f(x^\star(t),t)=0 
\qquad
\text{and}
\qquad
a^\star(t) = \dpar fx(x^\star(t),t) \leqs -a_0
\end{equation}
for all $t$, where $a_0$ is a positive constant. In the one-dimensional
case, $f$ always derives from a potential $V$, and $-a^\star(t)$ represents
the curvature at the bottom $x^\star(t)$ of a potential well. 

In the deterministic case $\sigma=0$, solutions of~\eqref{st1} track the
equilibrium branch $x^\star(t)$ adiabatically. In fact, Tihonov's Theorem
\cite{Tihonov,Grad} asserts that for $\sigma=0$,~\eqref{st1} admits a
particular solution $\xbdet_t$ with an asymptotic expansion of the form 
\begin{equation}
\label{st3}
\xbdet_t = x^\star(t) + \eps \frac{\dot x^\star(t)}{a^\star(t)} +
\Order{\eps^2}.
\end{equation}
Since $a^\star(t)$ is negative, $\xbdet_t$ lies a little bit to the left of
$x^\star(t)$ if $x^\star(t)$ moves to the right, and vice versa. The
\defwd{adiabatic solution} $\xbdet_t$ attracts nearby solutions
exponentially fast in $t/\eps$. 

\begin{figure}
 \centerline{\psfig{figure=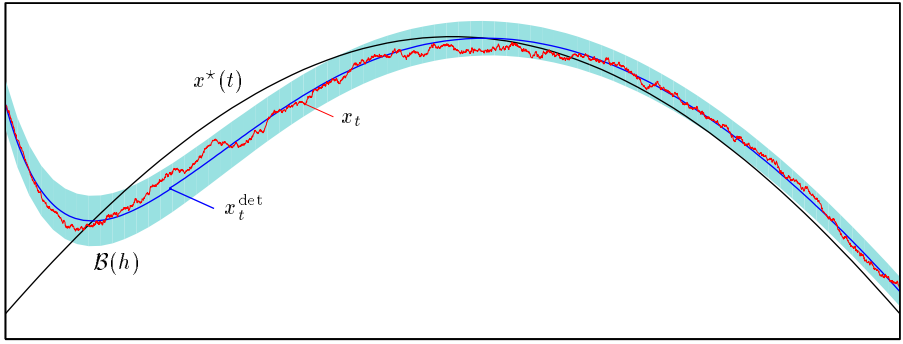,width=120mm,clip=t}}
 \captionspace
 \caption[]
 {A sample path of the SDE~\eqref{st1}, for
 $f(x,t)=a^\star(t)(x-x^\star(t))$  deriving from a quadratic single-well
 potential, and $g(t)\equiv1$. The potential well is located at
 $x^\star(t)=\sin(2\pi t)$, and has curvature $-a^\star(t)=4-2\sin(4\pi
 t)$. Parameter values are $\eps=0.04$ and $\sigma=0.025$. After a short
 transient motion, the deterministic solution $\xdet_t$ tracks $x^\star(t)$
 at a distance of order $\eps$. The path $x_t$ is likely to stay in the
 shaded set $\cB(h)$ (shown here for $h=3$), which is centred at $\xdet_t$
 and has time-dependent width of order $h\sigma/\sqrt{\abs{a^\star(t)}}$.}
\label{fig01}
\end{figure}

Consider now the SDE~\eqref{st1} with positive noise intensity. For the
sake of brevity, we assume that $g$ is positive and bounded away from zero.
In a nutshell, our main result can be formulated as follows: Up to Kramer's
time, paths starting near $\xbdet_0$ are concentrated in a \nbh\ of order
$\sigma g(t)/\sqrt{\abs{a^\star(t)}}$ of the deterministic solution with
the same initial condition, as shown in~\figref{fig01}. Larger noise
intensities and smaller curvatures thus lead to a larger spreading of
paths. This result holds as long as the spreading is smaller than the
distance between $x^\star(t)$ and the nearest unstable equilibrium
(i.\,e., the nearest saddle of the potential). 

To make this claim mathematically precise, we need a few definitions. For
simplicity, we discuss first the particular case $x_0=\xbdet_0$. We
use the notations
\begin{equation}
\label{st4}
a(t) = \dpar fx (\xbdet_t,t), 
\qquad
\alpha(t,s) = \int_s^t a(u)\6u
\qquad
\text{and} 
\qquad
\alpha(t)=\alpha(t,0).
\end{equation}
Note that by~\eqref{st3}, $a(t)=a^\star(t)+\Order{\eps}$ is negative for
sufficiently small $\eps$. The main idea is that $x_t-\xbdet_t$ is well
approximated by a generalized Ornstein--Uhlenbeck process, with
time-dependent damping $a(t)/\eps$ and diffusion coefficient $\sigma
g(t)/\sqrt\eps$. This process is obtained by linearizing the
SDE~\eqref{st1} around $\xbdet_t$, and has variance
\begin{equation}
\label{st5}
v(t) = \frac{\sigma^2}\eps \int_0^t \e^{2\alpha(t,s)/\eps} g(s)^2\6s. 
\end{equation}
The function $v(t)$ solves the ordinary differential equation (ODE)
$\eps\dot v = 2a(t)v + \sigma^2 g(t)^2$. In analogy with~\eqref{st3}, this
equation also admits a particular solution $\bar v(t)$ satisfying
\begin{equation}
\label{st6}
\bar v(t) = \frac{\sigma^2}{2\abs{a(t)}} \bigbrak{g(t)^2+\Order{\eps}},
\end{equation} 
and since $\alpha(t)\leqs-a_0t$ for $t\geqs0$, the variance
$v(t)=\bar v(t)-\bar v(0)\e^{2\alpha(t)/\eps}$ approaches $\bar v(t)$
exponentially fast. 
We now introduce the set 
\begin{equation}
\label{st7}
\cB(h) = \bigsetsuch{(x,t)}{\abs{x-\xbdet_t} < h\sqrt{\bar v(t)}}, 
\end{equation}
which depends on a real parameter $h>0$. The strip $\cB(h)$ is centred in
the adiabatic solution $\xbdet_t$ tracking the bottom of the potential
well, and has time-dependent width $h\sigma
g(t)/\sqrt{2\abs{a^\star(t)}}\brak{1+\Order{\eps}}$. To lowest order in
$\eps$ and $\sigma$, $\cB(h)$ coincides with the points in the potential
well for which $V(x,t)-V(\xbdet_t,t)$ is smaller than $(\frac12h\sigma
g(t))^2$. 

The main result is that for $h\gg 1$, paths $\set{x_s}_{s\geqs0}$ are
unlikely to leave the set $\cB(h)$ before Kramers' time. Equivalently, the
\defwd{first-exit time} 
\begin{equation}
\label{st8}
\tau_{\cB(h)} = \inf\bigsetsuch{t>0}{(x_t,t) \not\in \cB(h)} 
\end{equation}
is unlikely to be smaller than $T_{\math{Kramers}}$. Indeed, one can prove
the following estimate (see \cite[Theorem~2.4]{BG1} and
\cite[Theorem~2.2]{BG4}). There is a constant $h_0>0$ such that 
\begin{equation}
\label{st9}
\bigprob{\tau_{\cB(h)}>t} \leqs C(t,\eps) \e^{-\kappa h^2}
\end{equation}
holds for all $t>0$ and all $h\leqs h_0/\sigma$, where 
\begin{equation}
\label{st10}
C(t,\eps) = \frac{\abs{\alpha(t)}}{\eps^2} + 2
\qquad
\text{and}
\qquad
\kappa = \frac12 - \Order{\eps} - \Order{\sigma h}.
\end{equation}
The exponential term $\e^{-\kappa h^2}$ in~\eqref{st9} is independent of
time, and becomes small as soon as $h\gg1$. The constant $h_0$ depends
on $f$ and is the smaller the smaller $a_0$ is: The flatter the well, 
the more restrictive the condition $h\leqs h_0/\sigma$ becomes. 
The prefactor $C(t,\eps)$, which grows as time increases (and is
certainly not optimal) only leads to subexponential corrections on the
time scale $T_{\math{forcing}}$. Some time dependence of the prefactor is
to be expected, as it reflects the fact that occasionally a path will
make an unusually large excursion, and the longer we wait the more
excursions we will observe. The prefactor also depends on $\eps$. A
factor $1/\eps$ is due to the fact that we are working on the time
scale $T_{\math{forcing}}$, while the actual factor of $1/\eps^2$
allows us to obtain the best possible exponent. Choosing $\kappa$
slightly smaller allows to replace $\eps^2$ by the more natural $\eps$
in the definition of $C(t,\eps)$. Thus we find that paths are unlikely
to leave $\cB(h)$ before time $t$, provided $h_0^2/\sigma^2 \geqs h^2
\gg \log C(t,\eps)$.  

The same results hold if $x_t$ does not start on the adiabatic
solution $\xbdet_t$, but in some deterministic $x_0$ sufficiently
close to it. Then $\xbdet_t$ has to be replaced in~\eqref{st4} and
\eqref{st7} by the solution $\xdet_t$ of the deterministic equation
with initial condition $x_0$. We still have that $a(t)$ is negative
(and bounded away from zero), but note that~\eqref{st6} may not hold for
very small $t$, when $\xdet_t$ has not yet approached $\xbdet_t$.  

If the potential $V$ grows at least quadratically for large $\abs{x}$, one
can deduce from~\eqref{st9} that the moments of $\abs{x_t-\xdet_t}$ are
bounded by those of a centred Gaussian distribution with variance of order
$\bar v(t)$, for times small compared to Kramers' time
\cite[Corollary~2.4]{BG4}, even if $V$ has other potential wells than the
one at $x^\star(t)$. Assume for instance that $V$ has two potential wells,
with the shallower one at $x^\star(t)$. Then the system is in metastable
\lq\lq equilibrium\rq\rq\ for an exponentially long time span during
which the existence of the deeper well is not felt. 

Similar statements are valid in the multidimensional case (in which $f$
does not necessarily derive from a potential). Let $x^\star(t)$ be an
equilibrium branch of $f$, and denote by $A^\star(t)$ the Jacobian matrix of
$f$ at $x^\star(t)$. We assume that the eigenvalues of $A^\star(t)$ have
real parts smaller than some negative constant $-a_0$ for all times, so that
$x^\star(t)$ is asymptotically stable. In the deterministic case $\sigma=0$,
Tihonov's theorem shows the existence of an adiabatic solution 
\begin{equation}
\label{st11}
\xbdet_t = x^\star(t) + \eps A^\star(t)^{-1} \dot x^\star(t) +
\Order{\eps^2},
\end{equation}
which attracts nearby orbits exponentially fast. Let $A(t)$ be the
Jacobian matrix of $f$ at $\xbdet_t$. It satisfies
$A(t)=A^\star(t)+\Order{\eps}$. The solution of the SDE~\eqref{in1}
linearized at $\xbdet_t$ has a Gaussian distribution, with covariance
matrix  
\begin{equation}
\label{st12}
X(t) = \frac{\sigma^2}\eps \int_0^t U(t,s)G(s)G(s)^TU(t,s)^T \6s,
\end{equation}
where $U(t,s)$ is the fundamental solution of $\eps\dot y=A(t)y$ with
initial condition $U(s,s)=\one$. To keep the presentation simple, we
will assume that the smallest eigenvalue of $G(s)G(s)^T$ is bounded away
from zero and the largest one is bounded above. Note that $X(t)$ obeys
the ODE  $\eps\dot X = AX + XA^T + \sigma^2 GG^T$, and approaches
exponentially fast a matrix $\Xbar(t)$ which satisfies
\begin{equation}
\label{st13}
\Xbar(t) = \Xbar_0(t) + \Order{\eps},
\qquad
\text{where }
A\Xbar_0 + \Xbar_0 A^T = -\sigma^2 GG^T.
\end{equation}
Given a deterministic solution $\xdet_t$, the definition of the set
$\cB(h)$ reads now  
\begin{equation}
\label{st14}
\cB(h) = \bigsetsuch{(x,t)}{(x-\xdet_t)^T \Xbar(t)^{-1} (x-\xdet_t) < h^2},
\end{equation}
and~\eqref{st9} generalizes to the following statement 
(see~\cite[Theorem~6.1]{BG4} for a discussion; the proof will be given
in~\cite{BG6}): There is a constant $h_0>0$ such that for all
$h\leqs h_0/\sigma$ and all $\kappa\in(0,1/2)$, 
\begin{equation}
\label{st15}
\bigprob{\tau_{\cB(h)}>t} \leqs 
C(t,\eps) \e^{-\kappa h^2(1-\Order{\eps}-\Order{\sigma h})},
\end{equation}
where
\begin{equation}
\label{st16}
C(t,\eps) = \Bigpar{\frac t{\eps^2}+1} \Bigpar{\frac1{1-2\kappa}}^{n/2},
\end{equation}
$n$ being the dimension of $x$. 
Paths are thus concentrated, up to a given time $t$, in sets of the form
$\cB(h)$, which have an ellipso{\"\i}dal cross-section defined by
$\Xbar(t)$. Again the parameter $h$ must satisfy $h_0^2/\sigma^2 \geqs
h^2 \gg \log C(t,\eps)$. 
  
This result can be used, in particular, to understand the effect of
coloured noise. Assume for instance that the one-dimensional system 
\begin{equation}
\label{st17}
\6x_t = f(x_t,\eps t)\6t + g(\eps t)\6Z_t
\end{equation}
is not driven by white noise, but by an Ornstein--Uhlenbeck process $Z_t$
obeying the SDE
\begin{equation}
\label{st18}
\6Z_t = -\gamma Z_t \6t + \sigma\6W_t.
\end{equation}
The equations~\eqref{st17} and~\eqref{st18} can be rewritten, on the time
scale $1/\eps$, as a two-dimensional system of the form~\eqref{in1}
for $(x_t,Z_t)$. We assume that $f$ has a stable equilibrium branch
$x^\star(t)$ with linearization $a^\star(t)\leqs-a_0<0$. To leading
order in $\eps$, the asymptotic covariance matrix~\eqref{st13} is given by 
\begin{equation}
\label{st19}
\Xbar_0(t) = \sigma^2 
\begin{pmatrix}
\vrule height 11pt depth 16pt width 0pt
\dfrac{g(t)^2}{2(\gamma+\abs{a^\star(t)})} & 
\dfrac{g(t)}{2(\gamma+\abs{a^\star(t)})} \\
\vrule height 16pt depth 11pt width 0pt
\dfrac{g(t)}{2(\gamma+\abs{a^\star(t)})} &
\dfrac1{2\gamma}
\end{pmatrix}.
\end{equation}
The conditions on $GG^T$ mentioned above can be relaxed
(c.\,f.~\cite[Theorem~6.1]{BG4}), so that~\eqref{st15} is
applicable. We find in particular that the path $\set{x_t}_{t\geqs0}$
is concentrated in a strip of width proportional to $\sigma
g(t)/\sqrt{\gamma+\abs{a^\star(t)}}$, centred around $\xdet_t$. Hence
larger \lq\lq noise colour\rq\rq\ $\gamma$ yields a smaller spreading
of the paths, in the same way as if the curvature of the potential
were increased by $\gamma$.  


\section{Stochastic resonance}
\label{sec_sr}

In the previous section, we have seen that on a certain time scale, paths
typically remain in metastable equilibrium. With overwhelming
probability, they are concentrated in a strip of order $\sigma
g(t)/\sqrt{\abs{a^\star(t)}}$ near the bottom of a potential well with 
curvature $\abs{a^\star(t)}$. This roughly holds as long as the strip does
not extend to the nearest saddle of the potential. New phenomena may 
occur when this hypothesis is violated, either because the noise coefficient
$\sigma g(t)$ becomes too large, or because the curvature or the distance to
the saddle become too small. Then paths may overcome the potential barrier
and reach another potential well. This mechanism has various
interesting consequences, one of them being the effect called stochastic
resonance. 

Stochastic resonance (SR) was initially introduced as a possible
explanation for the close-to-periodic appearance of the major Ice Ages
\cite{BPSV}. While this explanation remains controversial, SR has been
detected in several other physical and biological systems, see for instance
\cite{MW,WM} for a review. 

The original model in~\cite{BPSV} is based on an energy balance of the
Earth in integrated form. The evolution of the mean surface temperature $T$
is described by the differential equation
\begin{equation}
\label{sr1}
c \dtot Tt = Q (1+A\cos\w t) (1-\alpha(T)) - E(T).
\end{equation}
Here the term $R_{\math{in}}=Q (1+A\cos\w t)$ is the incoming solar
radiation, where $Q$ denotes the solar constant, and the periodic term
models the 
effect of the Earth's varying orbital eccentricity. The amplitude $A$ of
this modulation is very small, of the order $5\times10^{-4}$, while its
period $2\pi/\w$ equals $92\,000$ years. The outgoing radiation
$R_{\math{out}} = \alpha(T)R_{\math{in}} + E(T)$ depends on the albedo
$\alpha(T)$ of the Earth and its emissivity. $c$ denotes the heat capacity.  

To account for the existence of two stable climate states (warm climate and
Ice Age), the right-hand side of~\eqref{sr1} should have two stable and one
unstable equilibrium points. The authors of~\cite{BPSV} postulate that
\begin{equation}
\label{sr2}
\gamma(T) = \frac Q{E(T)} (1-\alpha(T)) - 1 
= \beta \biggpar{1-\frac T{T_1}} \biggpar{1-\frac T{T_2}} 
\biggpar{1-\frac T{T_3}}, 
\end{equation}
where $T_1=278.6$ K and $T_3=288.6$ K are the representative temperatures of
the two stable states, and $T_2=283.3$ K represents the unstable state.
Since $E(T)\sim T^4$ varies little on this range, the problem can be further
simplified by neglecting the $T$-dependence of $E(T)\simeq\langle E\rangle$.
Equation~\eqref{sr1} becomes
\begin{equation}
\label{sr3}
\dtot Tt = \frac{\langle E\rangle}c \biggbrak{\beta \biggpar{1-\frac T{T_1}} 
\biggpar{1-\frac T{T_2}} \biggpar{1-\frac T{T_3}} (1+A\cos\w t) + 
A\cos\w t}. 
\end{equation}
The parameter $\beta$ is related to the relaxation time $\tau\simeq8$ years
of the system via 
\begin{equation}
\label{sr4}
\frac 1\tau = \frac{\langle E\rangle}c \beta \frac1{T_3} 
\biggpar{1-\frac {T_3}{T_1}} \biggpar{1-\frac {T_3}{T_2}}. 
\end{equation}
Let us now transform this system to a dimensionless form. We do this in two
steps: First we scale time by a factor $\w/2\pi$, so that in the new
variables, the system has period $1$. Then we introduce the variable
$x=(T-T_2)/\Delta T$, where $\Delta T=(T_3-T_1)/2 = 5$ K. The resulting
system is
\begin{equation}
\label{sr5}
\dtot xt = \frac1\eps \bigbrak{-x(x-x_1)(x-x_3)(1+A\cos2\pi t) 
+ K\cos2\pi t},
\end{equation}
where $x_1=(T_1-T_2)/\Delta T \simeq -0.94$ and $x_3=(T_3-T_2)/\Delta T
\simeq 1.06$. The adiabatic parameter $\eps$ is given by
\begin{equation}
\label{sr6}
\eps = \frac{\w\tau}{2\pi} \,\frac{2(T_3-T_2)}{\Delta T} 
\simeq 1.8\times 10^{-4}.
\end{equation}
This confirms that we are in the adiabatic regime. Using the value 
$\langle E\rangle/c = 8.77\times 10^{-3}/4000$ $\text{Ks}^{-1}$ from
\cite{BPSV}, we find a driving amplitude 
\begin{equation}
\label{sr7}
K = \frac A\beta \frac{T_1T_2T_3}{(\Delta T)^3} \simeq 0.12.
\end{equation}
The term in brackets in~\eqref{sr5} derives from a double-well potential,
which is almost of the Ginzburg--Landau type~\eqref{in8}. If we set, for
simplicity, $x_1=-1$ and $x_3=1$, and neglect the term $A\cos2\pi t$, then
we obtain indeed a force deriving from the potential~\eqref{in8}, with
$\mu=1$ and $\lambda=K\cos2\pi t$. This potential has two wells if and only
if $\abs\lambda<\lc=2/3\sqrt3\simeq0.38$, and thus the amplitude $K$ of
the forcing is too small to enable transitions between the potential
wells. Note, however, that although $A$ is very small, $K$ is not negligible
compared to $\lc$. 

The main new idea in~\cite{BPSV} is that if one models the effect of the
\lq\lq weather\rq\rq\ by an additive noise term, then transitions between
potential wells not only become possible but, due to the periodic
forcing, these transitions will be more likely at some times than at
others, so that the evolution of $T$ can be close to periodic. We will
illustrate this on the model SDE
\begin{equation}
\label{sr8}
\6x_t = \frac1\eps \bigbrak{x_t-x_t^3+K\cos2\pi t}\6t +
\frac\sigma{\sqrt\eps} \6W_t.
\end{equation}
However, the results in~\cite{BG2} apply to a more general class of
periodically forced double-well potentials, including~\eqref{sr5}. 

\begin{figure}
 \centerline{\psfig{figure=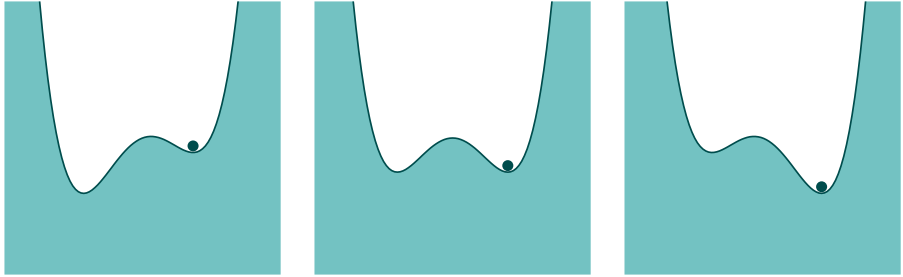,width=120mm,clip=t}}
 \captionspace
 \caption[]
 {The potential $V(x,t)=\frac14 x^4 - \frac12 x^2 - K\cos(2\pi t)x$, from
 which derives the drift term in~\eqref{sr8}. For $\cos(2\pi t)=0$, the
 potential is symmetric (middle), for integer times, the left-hand well
 approaches the saddle (right), while for half-integer times, the right-hand
 well approaches the saddle (left). If the amplitude $K$ is smaller than the
 threshold $\lc$, there is always a potential barrier, which an overdamped
 particle cannot overcome in the deterministic case. Sufficiently strong
 noise, however, helps the particle to switch from the shallower to the deeper
 well. This effect is the stronger the lower the barrier is, so that
 switching typically occurs close to the instants of minimal barrier height.}
\label{fig02}
\end{figure}

Various characterizations of the effect of noise on the dynamics of
\eqref{sr8}, and various measures of periodicity have been proposed. A
widespread approach uses the signal-to-noise ratio, a property of the power
spectrum of $x_t$, which shows peaks near multiples of the driving frequency
\cite{Fox,McNW,JH1}. For small driving amplitudes $K$, the signal-to-noise
ratio behaves like $\e^{-H/\sigma^2}/\sigma^4$, where $H$ is the height
of the potential barrier in the absence of periodic driving 
(i.\,e., for $K=0$). The signal's \lq\lq periodicity\rq\rq\ is thus
optimal for $\sigma^2=H/2$. 
A different approach is used in~\cite{Freidlin1}, where the
$L^p$-distance between sample paths and a periodic limiting function
is shown to converge to zero in probability as $\sigma\to0$. This result
requires $\eps$ to be of order $\e^{-2H/\sigma^2}$, which implies 
exponentially long forcing periods. 

We examine here a different regime, in which the forcing amplitude $K$ is
not necessarily a small parameter, but may approach $\lc$. In this way,
transitions become possible for values of $\eps$ which are not
exponentially small. The potential barrier is lowest at integer and
half-integer times. At integer times, the left-hand well approaches the
saddle, while at half-integer times, the right-hand well approaches the
saddle, c.\,f.~\figref{fig02}.

The minimal values $H_{\math{min}}$, $c_{\math{min}}$ and
$\delta_{\math{min}}$ of the barrier height, the curvature at the bottom of
the wells, and the distance between the bottom of one of the wells and
the saddle can be expressed as 
functions of a parameter $a_0=\lc-K$. For small $a_0$, they behave like
$H_{\math{min}}(a_0)\asymp \smash{a_0^{3/2}}$, $c_{\math{min}}(a_0)\asymp
\smash{a_0^{1/2}}$ and $\delta_{\math{min}}(a_0)\asymp \smash{a_0^{1/2}}$
(meaning $c_- \smash{a_0^{3/2}} \leqs H_{\math{min}}(a_0) \leqs c_+
\smash{a_0^{3/2}}$ for some positive constants $c_\pm$ independent of
$a_0$, and so on). 

Intuitively, our results from Section~\ref{sec_st} indicate that the
maximal spreading of paths is of order
$\sigma/\smash{c_{\math{min}}(a_0)^{1/2}}$, provided this value is smaller
than $\delta_{\math{min}}(a_0)$, i.\,e., provided $\sigma \ll
\smash{a_0^{3/4}}$. Assume for instance that we start at time $1/4$ (when
the potential is symmetric) near the right-hand potential well. We call
\defwd{transition probability} the probability $P_{\math{trans}}$ of
having reached the left-hand potential well by time $3/4$, after
passing through the configuration with the shallowest right-hand well. 
Extrapolating~\eqref{st9} with $h$ of the order
\smash{$\delta_{\math{min}}c_{\math{min}}^{1/2}/\sigma \asymp 
H_{\math{min}}^{1/2}/\sigma$}, we find
\begin{equation}
\label{sr9}
P_{\math{trans}} \leqs \frac {\text{\it const}}{\eps^2} 
\e^{-\text{\it const } a_0^{3/2}/\sigma^2}
= \frac {\text{\it const}}{\eps^2} 
\e^{-\text{\it const } H_{\math{min}}/\sigma^2}
\qquad\qquad
\text{for $\sigma\leqs a_0^{3/4}$.}
\end{equation}
Note the similarity with Kramers' time for the potential frozen at the
moment of minimal barrier height.

A bound of this form can indeed be proved, but~\eqref{sr9} turns
out to be a little bit too pessimistic for very small $a_0$.
This is a rather subtle dynamical effect, related to the behaviour
of the deterministic system. Recall that the set $\cB(h)$ in~\eqref{st7} is
defined via the linearization at the adiabatic solution $\xbdet_t$,
not at the
bottom $x^\star(t)$ of the potential well. This distinction is irrelevant
as long as the minimal curvature remains of order one, but {\em not\/}
when it is a small parameter. In that case, the asymptotic expansion
\eqref{st3} does not necessarily converge. Using methods from singular
perturbation theory~\cite{BK2}, one can show that $\xbdet_t$ never
approaches the saddle closer than a distance of order $\sqrt\eps$, so that
the curvature at $\xbdet_t$ never becomes smaller than a quantity of order
$\sqrt\eps$, even if $a_0<\eps$. As a consequence, for $a_0<\eps$, the
system behaves as if there were an effective potential barrier of
height $\eps^{3/2}$.  

In fact, one can prove the following bound (see~\cite[Theorem~2.6]{BG2} and
\cite[Theorem~3.1]{BG4}): There exist constants $C,\kappa>0$ such that 
\begin{equation}
\label{sr10}
P_{\math{trans}} \leqs \frac {C}{\eps} \e^{-\kappa \sigmac^2/\sigma^2}
\qquad\qquad
\text{for $\sigma\leqs\sigmac = (a_0\vee\eps)^{3/4}$,}
\end{equation}
where $a\vee b$ denotes the maximum of two real numbers $a$ and $b$. In
addition, paths remain concentrated in a set $\cB(h)$ of the form
\eqref{st7}. Examining the behaviour of the integral~\eqref{st5}, one can
show that the width of $\cB(h)$ behaves, near $t=1/2$, like
$h\sigma/(\abs{t-1/2}^{1/2}\vee\smash{\sigmac^{1/3}})$. The various
exponents entering these relations do not depend on the details of the
potential, but only on some qualitative properties of the \lq\lq avoided
bifurcation\rq\rq, and can be deduced geometrically from a Newton polygon
\cite{BK2}. 

\begin{figure}
 \centerline{\psfig{figure=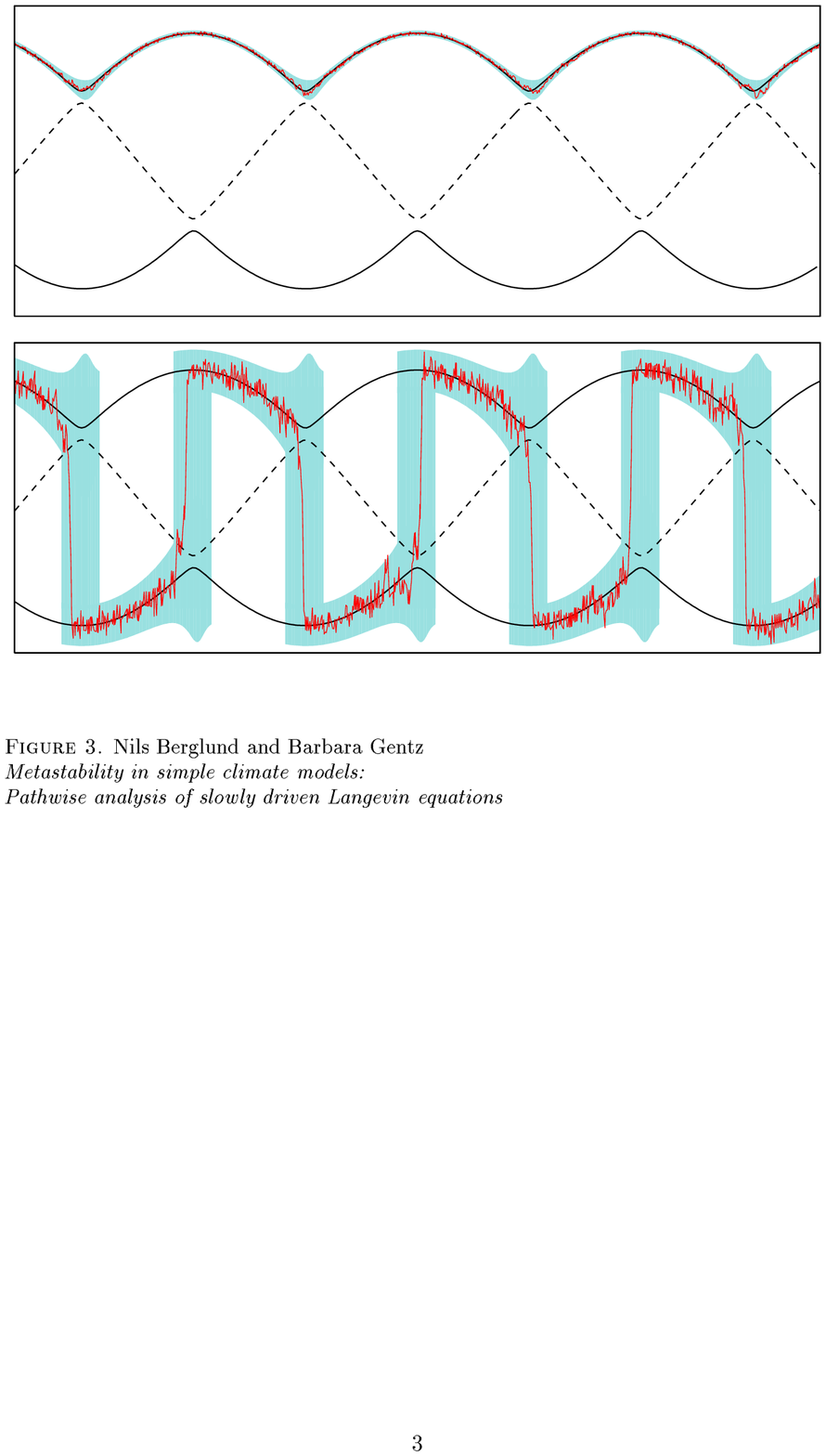,width=120mm,clip=t}}
 \captionspace
 \caption[]
 {Sample paths of the SDE~\eqref{sr8} for $\eps=a_0=0.005$, and
 $\sigma=0.02$ (upper picture) and $\sigma=0.14$ (lower picture). Full
 curves represent the location of potential wells, the broken curve
 represents the saddle. For weak noise, the path $x_t$ is likely to stay in
 the shaded set $\cB(h)$, centred at the deterministic solution tracking the
 right-hand well. The maximal width of $\cB(h)$ is of order
 $h\sigma/(a_0\vee\eps)^{1/4}$ and is reached at half-integer times. For strong
 noise, typical paths stay in the shaded set which switches back and forth
 between the wells at integer and half-integer times. The width of the
 vertical strips is of order $\sigma^{2/3}$. The \lq\lq bumps\rq\rq\ are due
 to the fact that one of the wells becomes very flat during the transition
 window so that paths might also make excursions away from the saddle.}
\label{fig03}
\end{figure}

What happens when $\sigma$ exceeds the threshold value $\sigmac$? Away from
half-integer times, the right-hand well may still be sufficiently deep to
confine the paths. However, there are time intervals near half-integer $t$
during which it becomes possible to overcome the barrier. Near $t=1/2$, the
curvature $c(t)$ at $\xbdet_t$ and the distance between $\xbdet_t$ and the
saddle both behave like $\abs{t-1/2}\vee\smash{\sigmac^{2/3}}$. Transitions
thus become possible for $\abs{t-1/2}\leqs\sigma^{2/3}$. 

During this time interval, the process $x_t$ makes a certain number of
attempts to overcome the barrier. If the saddle is reached, $x_t$ has
roughly equal probability to fall back into the right-hand well, in which
case it will make further attempts to cross the barrier, or to fall into the
deeper left-hand well, where it is likely to stay during the next
half-period. One can show that the typical time for each excursion is of
order $\eps/c(t)$. Although the different attempts are not independent, the
probability {\em not\/} to reach the left-hand well during the transition
window $\abs{t-1/2}\leqs\sigma^{2/3}$ behaves roughly like $(1/2)^N$, where
$N$ is the maximal number of possible excursions. 

These arguments can be used to show (see~\cite[Theorem~2.7]{BG2} and
\cite[Theorem~3.1]{BG4}) that there exist constants $C,\kappa>0$ such that 
\begin{equation}
\label{sr11}
P_{\math{trans}} \geqs 1 - C \e^{-\kappa \sigma^{4/3}/(\eps\abs{\log\sigma})}
\qquad\qquad
\text{for $\sigma\geqs\sigmac$.}
\end{equation}
The factor $\sigma^{4/3}$ is proportional to the integral of $c(t)$ over
the transition window, and the factor $\abs{\log\sigma}$ takes into account
the time needed to travel from the saddle to the left-hand well.
Amplification by SR is thus optimal for noise intensities just above the
threshold $\sigmac$, because stronger noise intensities will gradually
blur the signal.

In the large-noise regime $\sigma \geqs \sigmac$, the vast majority of
paths stay in a strip switching back and forth between potential wells
each time the barrier height becomes minimal, as shown in~\figref{fig03}. 

Paths spend approximately half the time (for 
$1/4<t<1/2$, $3/4<t<1$, and so on) in metastable equilibrium in the
shallower potential well. This differs from the quasistatic picture, when the
driving period is larger than the maximal Kramers time, and paths spend most
of the time in the deeper potential well with occasional excursions to the
shallower one. 

While the details of the transition process depend on the potential, the
exponents in~\eqref{sr10} and~\eqref{sr11} depend only on qualitative
properties of the avoided bifurcation. Other exponents arise, for instance,
if $V$ is a symmetric potential with modulated barrier height of the form
\eqref{in8} with $\lambda=0$ and $\mu(t)=a_0+1-\cos2\pi t$,
c.\,f.~\cite[Theorem~3.2]{BG4}. Here an additional feature can be observed:
For sufficiently strong noise, the process is likely to reach the saddle
during a certain transition window, but due to symmetry, it has about equal
probability to be in either of the wells when transitions become
unlikely again. Observing the process for several periods, we see that near
the instants of minimal barrier height, the process chooses  randomly
between potential wells, with probability exponentially close to $1/2$ for
choosing either.  

One can also consider the effect of coloured noise on SR. If the system is
driven by an Ornstein--Uhlenbeck process with damping $\gamma$, the typical
spreading of paths will be smaller, making transitions more difficult. One
can show that transitions only become likely above a threshold noise
intensity $\sigmac$, given by 
\begin{equation}
\label{sr12}
\sigmac^2 =  (a_0\vee\eps)\bigpar{\gamma \vee (a_0 \vee \eps)^{1/2}}.
\end{equation}
If $\gamma<(a_0 \vee \eps)^{1/2}$, we recover the white-noise result, but
for larger $\gamma$, the threshold grows linearly with $\gamma$,
namely like $(a_0\vee\eps)\gamma$. 

It is, of course, not easy to decide whether the observed periodicity in
the appearance of Ice Ages can be explained by a simple, one-dimensional
SDE of the form~\eqref{sr8}. Our results show, however, that in order to
match the observations, the noise intensity should lie in a relatively
narrow interval. Too weak noise will not allow regular transitions between
stable states, while too strong noise increases the width of the
transition windows so much that although switching does occur, no
periodicity can be observed. 


\section{Hysteresis}
\label{sec_h}

The glacial cycle is not the only important bistable system in climate 
physics. Another wellknown example is the Atlantic thermohaline
circulation. At present time, the Gulf Stream transports enormous amounts
of heat from the Tropics as far north as the Barents Sea, causing the
current mild climate in Western Europe. It is believed, however, that this
has not always been the case in the past, and that during long time spans,
the thermohaline circulation was locked in a stable state with far less
heat transported to the North (see for instance~\cite{Rahmstorf}).

A simple model for oceanic circulation showing bistability is
Stommel's box model~\cite{Stommel}, where the ocean is represented by two
boxes, a low-latitude box with temperature $T_1$ and salinity $S_1$,
and a high-latitude box with temperature $T_2$ and salinity
$S_2$. Here we will follow the presentation in~\cite{Cessi}, where
the intrinsic dynamics of salinity and of temperature are not modeled in
the same way.
The differences $\Delta T=T_1-T_2$ and $\Delta S=S_1-S_2$ are assumed
to evolve according to the equations  
\begin{align}
\label{h1}
\dtot{}t \Delta T &= -\frac1{\tau_{\math{r}}} (\Delta T-\theta) 
- Q(\Delta \rho) \Delta T \\
\label{h2}
\dtot{}t \Delta S &= \frac{F(t)}H S_0  
- Q(\Delta \rho) \Delta S.
\end{align}
Here $\tau_{\math{r}}$ is the relaxation time of $\Delta T$ to its
reference value $\theta$, $S_0$ is a reference salinity, and $H$ is the
depth of the model ocean. $F(t)$ is the freshwater flux, modeling
imbalances between evaporation (which dominates at low latitudes) and
precipitation (which dominates at high latitudes). 
The dynamics of $\Delta T$ and $\Delta S$ are coupled via the density
difference $\Delta\rho$, approximated by the linearized equation of state
\begin{equation}
\label{h1.5}
\Delta\rho = \alpha_S\Delta S - \alpha_T\Delta T,
\end{equation}
which induces an exchange of mass $Q(\Delta\rho)$ between the
boxes. We will use here Cessi's model~\cite{Cessi} for $Q$,
\begin{equation}
\label{h3}
Q(\Delta\rho) = \frac1{\tau_{\math{d}}} + \frac qV \Delta\rho^2,
\end{equation}
where $\tau_{\math{d}}$ is the diffusion time scale, $q$ the Poiseuille
transport coefficient and $V$ the volume of the box. Stommel uses a
different relation, with $\Delta\rho^2$ replaced by $\abs{\Delta\rho}$, but
we will not make this choice here because it leads to a singularity. 

Using the dimensionless variables $y=\alpha_S \Delta
S/(\alpha_T\theta)$, $z=\Delta T/\theta$ and rescaling time by a
factor $\tau_{\math d}$, \eqref{h1} and \eqref{h2} can be rewritten as
\begin{equation}
\begin{split}
\label{h4}
\dot y &= p(t) - y \bigbrak{1+\eta^2(y-z)^2} \\
\eps_0 \dot z &= -(z-1) - \eps_0 z \bigbrak{1+\eta^2(y-z)^2},
\end{split}
\end{equation}
where $\eps_0=\tau_{\math{r}}/\tau_{\math{d}}$,
$\eta^2=\tau_{\math{d}}(\alpha_T\th)^2q/V$, and $p(t)$ is proportional to the
freshwater flux $F(t)$, with a factor $\alpha_S S_0\tau_{\math{d}}/
(\alpha_T\theta H)$. Cessi uses the estimates $\eta^2\simeq 7.5$,
$\tau_{\math{r}}\simeq 25$ days and $\tau_{\math d}\simeq 219$ years. This
yields $\eps_0\simeq 3\times10^{-4}$, implying that~\eqref{h4} is a
slow--fast system. Tihonov's theorem~\cite{Tihonov} allows us to reduce the
dynamics to the attracting slow manifold $z=1+\Order{\eps_0}$. 
To leading order, we thus find
\begin{equation}
\label{h5}
\dot y = -y \bigbrak{1+\eta^2(y-1)^2} + p(t). 
\end{equation} 

Stochasticity shows up in this model through the weather-dependent term
$p(t)$. To model long-scale variations in the typical weather, we will
assume that $p(t)$ can be represented as the sum of a periodic term $\bar
p(t)$ and white noise, where the period $1/\eps$ of  $\bar p(t)$ is much
longer than the diffusion time, which equals $1$. (Recall that we have
already rescaled time by a factor of $\tau_{\math{d}}$.) We thus obtain the
SDE 
\begin{equation}
\label{h6}
\6y_t = f(y_t,t)\6t + \sigma_0\6W_t,
\qquad
\text{where}
\qquad
f(y,t) = -y \bigbrak{1+\eta^2(y-1)^2} + \bar p(t).
\end{equation}
Note that $f$ has an inflection point at $y=2/3$, and that 
\begin{equation}
\label{h7}
\eta f\biggpar{\frac23 + \frac x\eta,t} = 
\eta \biggbrak{\bar p(t) - \frac23 - \frac 2{27}\eta^2} 
+ \biggbrak{\frac13\eta^2-1}x - x^3,
\end{equation}
which derives from the Ginzburg--Landau potential~\eqref{in8} with
parameters $\mu=(\eta^2/3-1)$ and $\lambda(t)=\eta(\bar p(t) - 2/3 -
2\eta^2/27)$. As we already know, the potential has two wells if and only
if $\lambda^2<\lc^2 = 4\mu^3/27$, which means, for $\eta^2=7.5$, that 
$\bar p \in[0.96,1.48]$.  The double-well potential is symmetric for $\bar
p=\bar p_0 = 2/3+2\eta^2/27\simeq1.22$.  

For a deterministic forcing given by $\lambda(t)=K\cos2\pi\eps t$, the SDE
for $x=\eta(y-2/3)$ becomes, on the time scale $1/\eps$, 
\begin{equation}
\label{h8}
\6x_t = \frac1\eps \bigbrak{\mu x - x^3 + K\cos2\pi t}\6t +
\frac{\sigma}{\sqrt\eps} \6W_t,
\end{equation}
where $\sigma=\sigma_0\eta$. This SDE is of the same form as~\eqref{sr8}.
While in Section~\ref{sec_sr}, we assumed $K<\lc$, we will now allow $K$ to
exceed $\lc$, so that the difference $a_0=K-\lc$ may change sign. (Note
that in Section~\ref{sec_sr}, $a_0$ had the opposite sign.) 

\begin{figure}
 \centerline{\psfig{figure=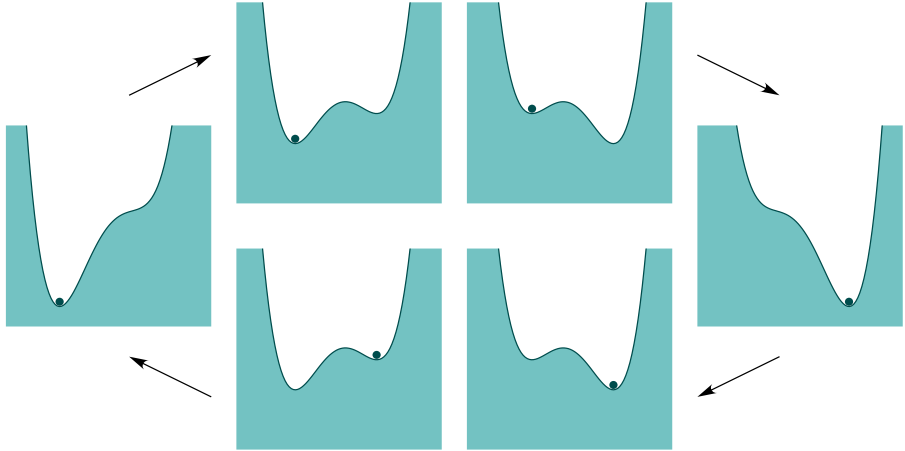,width=120mm,clip=t}}
 \captionspace
 \caption[]
 {The potential $V(x,t) = \frac14x^4 - \frac12x^2 - \lambda(t)x$, with
 $\lambda(t)=K\cos(2\pi t)$, when $K$ exceeds the threshold $\lc$. In the
 deterministic case, with $\eps\ll1$, the overdamped particle jumps to a new
 well whenever $\abs{\lambda(t)}$ becomes larger than $\lc$, leading to
 hysteresis. Larger values of $\eps$ increase the size of hysteresis cycles,
 but additive noise of sufficient intensity decreases the size of typical
 cycles, because it advances transitions to the deeper well.}
\label{fig04}
\end{figure}

In the deterministic case $\sigma=0$, Equation~\eqref{h8} has been used to
model a laser~\cite{JGRM}, and a similar equation describes the dynamics of
a mean-field Curie--Weiss ferromagnet~\cite{TO}. In the limit of infinitely
slow forcing, solutions always remain in the same potential well if $K<\lc$.
If $K>\lc$, however, the well tracked by $x_t$ disappears in a saddle--node
bifurcation when $\abs{\lambda(t)}$ crosses $\lc$ from below, causing $x_t$ to
jump to the other well, which leads to hysteresis, see~\figref{fig04}.

For positive $\eps$, the system does not react immediately to changes in the
potential, so that the hysteresis cycles are deformed. One can show
\cite{JGRM,BK2} that 
\begin{itemiz}
\item	For $K\leqs \lc+\Order{\eps}$, $x_t$ always tracks the same
potential well, at a distance at most of order $\eps/\sqrt{\abs{a_0}}$ if
$a_0\leqs-\eps$, and of order $\sqrt\eps$ if $\abs{a_0}$ is of order $\eps$.  
\item	For $K\geqs \lc+\Order{\eps}$, $x_t$ is attracted by a hysteresis
cycle, which is larger than the static hysteresis cycle; in particular,
$x_t$ crosses the $\lambda$-axis when $\lambda(t)=K\cos2\pi t=\lambda^0$,
where $\lambda^0$ satisfies
\begin{equation}
\label{h9}
\abs{\lambda^0} - \lc \asymp \eps^{2/3}a_0^{1/3},
\qquad
\text{with $a_0=K-\lc$.}
\end{equation}
\end{itemiz}

Additive noise will also influence the shape of hysteresis cycles, because
it can kick the state over the potential barrier, as has been noted
in~\cite{Monahan} in the context of the thermohaline circulation. For
positive noise intensities $\sigma$, the value $\lambda^0$ at which
$x_t$ crosses the $\lambda$-axis, becomes a random variable. 
Assume for instance that we start at time $t_0=1/4$ in 
the right-hand potential well. We define 
\begin{equation}
\label{h10}
\tau^0(\w) = \inf\Bigsetsuch{t\in\Bigbrak{\frac14,\frac34}}{x_t(\w)<0}, 
\qquad
\lambda^0(\w) = \lambda(\tau^0(\w)),
\end{equation} 
with the convention that $\tau^0(\w)=\infty$ and $\lambda^0(\w)=\infty$ if
$x_t(\w)>0$ for all $t\in[\frac14,\frac34]$. We thus have
$\tau^0\in[\frac14,\frac34]\cup\set\infty$ and
$\lambda^0\in[-K,K]\cup\set\infty$. We will indicate the
parameter-dependence by $\lambda^0=\lambda^0(\eps,\sigma)$, keeping in mind
that this random variable also depends on $a_0$ and $\mu$. 
In the deterministic case, $\lambda^0(\eps,0)=\infty$ if
$K\leqs\lc+\Order{\eps}$, and $\lambda^0(\eps,0)$ satisfies~\eqref{h9} if
$K\geqs\lc+\Order{\eps}$. 

As we know from the previous section, for $K<\lc$, there is an
amplitude-dependent threshold noise level $\sigmac$ such that during
one period, $x_t$ is unlikely to cross the potential barrier for
$\sigma\ll\sigmac$, while it is likely to cross it for $\sigma\gg\sigmac$.
In fact, in the latter case, there is a large probability to cross the
barrier a time of order $\sigma^{2/3}$ before the instant $t=1/2$ of
minimal barrier height, when $\lambda$ is of order $\lc-\sigma^{4/3}$. In
that case, the hysteresis cycle will be {\em smaller\/} than the static
cycle. A similar distinction between a small-noise and a large-noise regime
exists for large-amplitude forcing. 

It turns out that the distribution of $\lambda^0$ can be of three different
types, depending on the values of the parameters
(c.\,f.~\figref{fig05} and~\figref{fig06}):
\begin{itemiz}
\item	{\bf Case I -- Small-amplitude regime:}
$a_0\leqs\text{{\it const }}\eps$ and
$\sigma\leqs(\abs{a_0}\vee\eps)^{3/4}$.

Then $x_t$ is unlikely to cross the potential barrier, and there are
constants $C,\kappa>0$ such that (see~\cite[Theorem~2.3]{BG3})
\begin{equation}
\label{h11}
\bigprob{\lambda^0<\infty} \leqs \frac C\eps
\e^{-\kappa(\abs{a_0}\vee\eps)^{3/2}/\sigma^2}.
\end{equation}
The probability to observe a \lq\lq macroscopic\rq\rq\ hysteresis cycle is
very small, as most paths are concentrated in a small \nbh\ of the
bottom of the right-hand potential well (\figref{fig05}a).

\begin{figure}
 \centerline{\psfig{figure=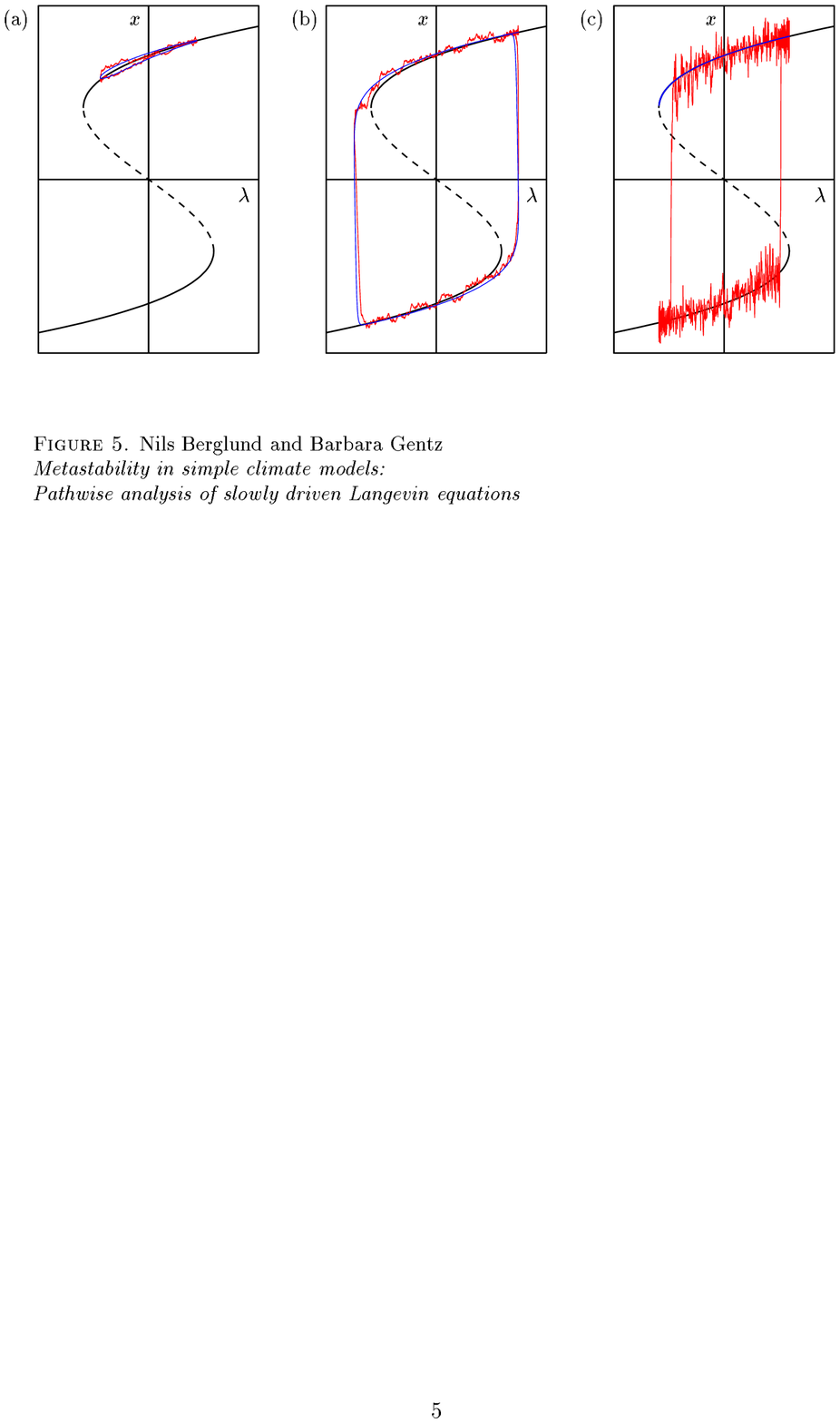,width=130mm,clip=t}}
 \captionspace
 \caption[]
 {Typical random hysteresis \lq\lq cycles\rq\rq\ in the three parameter
 regimes. (a) Case~I: Driving amplitude $K$ and noise intensity $\sigma$ are
 too small to allow the path to switch potential wells. (b) Case~II: For
 large amplitude but weak noise, the path tracks the deterministic
 hysteresis cycle, which is larger than the static one. (c) Case~III: For
 sufficiently strong noise, the path can overcome the potential barrier, so
 that typical hysteresis cycles are smaller than the static one.}
\label{fig05}
\end{figure}

\item	{\bf Case II -- Large-amplitude regime:}
$a_0\geqs\text{{\it const }}\eps$ and $\sigma\leqs\esa0^{1/2}$.

This regime is actually the most difficult to study, since the
deterministic solution jumps when $\abs{\lambda(t)}-\lc\asymp\esa0^{2/3}$, and
crosses a zone of instability before reaching the left-hand potential well.
One can show, however, that $\abs{\lambda^0}$ is concentrated in an
interval of length of order $\esa0^{2/3}$ around the deterministic value
\cite[Theorem~2.4]{BG3}. More precisely, there are constants $C,\kappa>0$
such that 
\begin{equation}
\label{h12}
\bigprob{\abs{\lambda^0}<\lc-L} \leqs \frac C\eps \e^{-\kappa
L^{3/2}/\sigma^2} 
\end{equation}
for $\esa0^{2/3}\leqs L\leqs L_0/\abs{\log\esa0}$, and 
\begin{equation}
\label{h13}
\bigprob{\abs{\lambda^0}<\lc+L_1\esa0^{2/3}} \leqs 
\frac C\eps \e^{-\kappa \eps\sqrt{a_0}/\sigma^2},
\end{equation}
where the constants $L_0, L_1>0$ are independent of the small parameters.
Hence it is unlikely to observe a substantially smaller value of
$\abs{\lambda^0}$ than the deterministic one, provided
$\sigma\ll\esa0^{1/2}$. On the other hand, there is a constant $L_2>L_1$
such that 
\begin{equation}
\label{h14}
\bigprob{\abs{\lambda^0}>\lc+L} \leqs 3 \e^{-\kappa
L/(\sigma^2\esa0^{2/3}\abs{\log\esa0})}
\end{equation}
for all $L\geqs L_2\esa0^{2/3}$. As a consequence, the vast majority of
hysteresis cycles will look very similar to the deterministic ones, which
are slightly {\em larger\/} than the static hysteresis cycle
(\figref{fig05}b). 

\begin{figure}
 \centerline{\psfig{figure=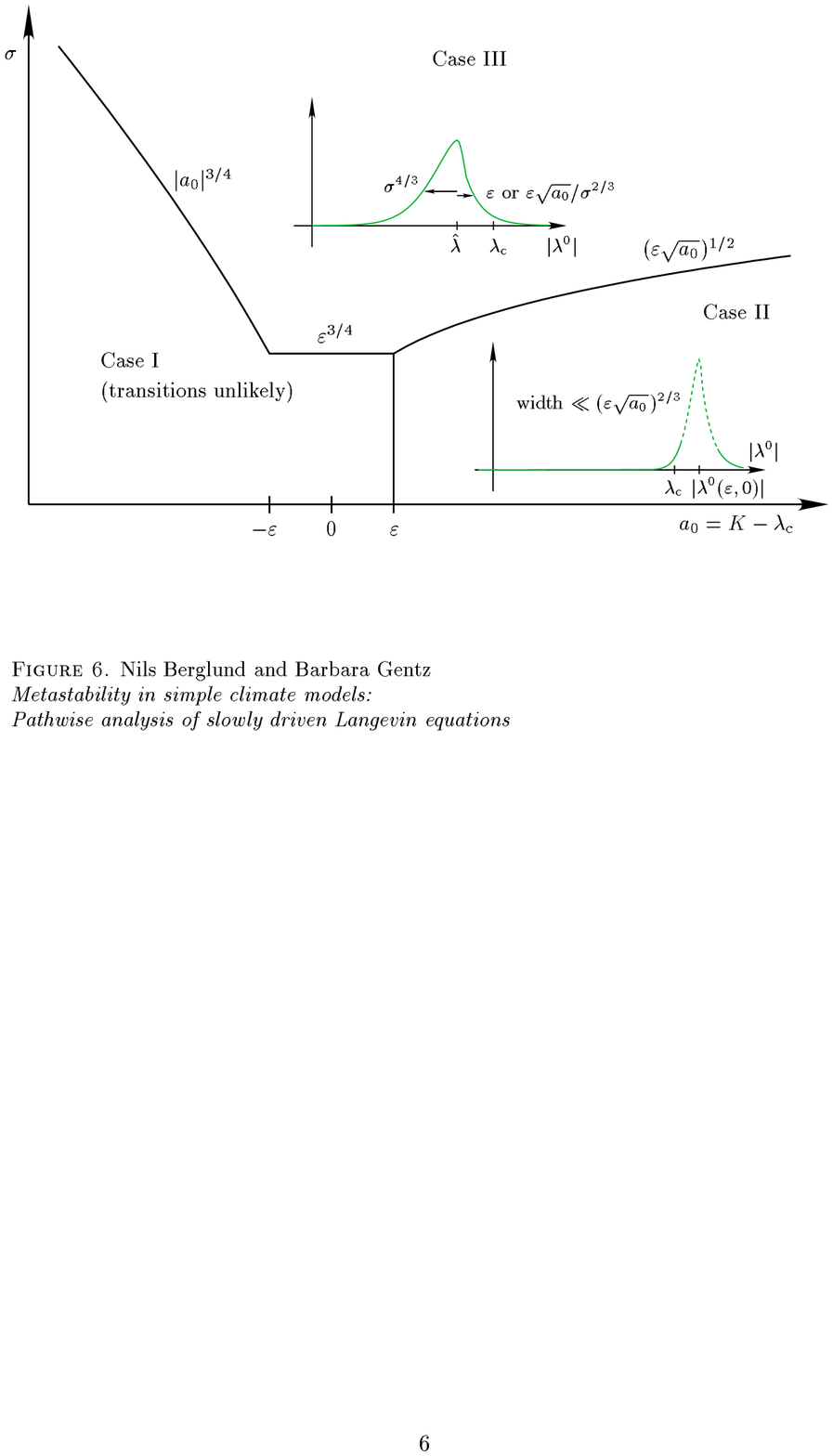,width=138mm,clip=t}}
 \captionspace
 \caption[]
 {The three hysteresis regimes, shown in the plane
 driving-amplitude--noise-intensity, for fixed driving frequency. The insets
 sketch the distribution of the random value $\lambda^0$ of the forcing
 $\lambda(t)$ when $x_t$ changes sign for the first time. In Case~I, such
 transitions are unlikely. In Case~II, $\abs{\lambda^0}$ is concentrated in
 an interval $[\lc+L_1\esa0^{2/3},\lc+L_2\esa0^{2/3}]$ containing the
 deterministic value $\abs{\lambda^0(\eps,0)}$. The broken curve indicates
 that we do not control the distribution inside this interval. In Case~III,
 $\abs{\lambda^0}$ is concentrated around a value $\hat\lambda$ which is
 smaller than $\lc$ by an amount of order $\sigma^{4/3}$. The distribution
 decays faster to the right, with a width of order $\eps$ (actually,
 $\eps\abs{\log\sigma}$) if $a_0\leqs\eps$ or $\sigma\geqs\abs{a_0}^{3/4}$,
 and of order $\eps\sqrt{a_0}/\sigma^{2/3}$ if $a_0\geqs\eps$ and
 $\sigma\leqs\abs{a_0}^{3/4}$.}
\label{fig06}
\end{figure}

\item	{\bf Case III -- Large-noise regime:}
Either $a_0\leqs\eps$ and $\sigma\geqs(\abs{a_0}\vee\eps)^{3/4}$ or 
$a_0\geqs\eps$ and $\sigma\geqs\esa0^{1/2}$.

In this case, the noise is sufficiently strong to drive $x_t$ over the
potential barrier, with large probability, some time before the barrier is
lowest or vanishes, leading to a {\em smaller\/} hysteresis cycle than in the
deterministic case (\figref{fig05}c). 
It turns out that $\abs{\lambda^0}$ is always
concentrated around a (deterministic) value $\hat\lambda$ satisfying
$\lc-\hat\lambda\asymp\sigma^{4/3}$. It follows from
\cite[Proposition~5.1]{BG3} that 
\begin{equation}
\label{h15}
\bigprob{\abs{\lambda^0}<\hat\lambda-L} \leqs 
\frac C\eps \e^{-\kappa L^{3/2}/\sigma^2} +
\frac32\e^{-\kappa\sigma^{4/3}/(\eps\abs{\log\sigma})} 
\end{equation}
for $0\leqs L\leqs\hat\lambda$ and 
\begin{equation}
\label{h16}
\bigprob{\abs{\lambda^0}>\hat\lambda+L} \leqs 
\frac32\e^{-\kappa L/(\eps\abs{\log\sigma})}
\end{equation}
for positive $L$ up to $K-\hat\lambda$ if $a_0\leqs\eps$. If
$a_0\geqs\eps$, the same bound holds for $L\leqs\lc-\hat\lambda$, while the
behaviour for larger $L$ is described by~\eqref{h14}. The estimates
\eqref{h15} and~\eqref{h16} hold if $a_0\leqs\eps$ or
$\sigma>\smash{a_0^{3/4}}$. In the other case, two exponents are modified:
$\sigma^{4/3}/(\eps\abs{\log\sigma})$ is replaced by
$\sigma^2/(\eps\sqrt{a_0}\mskip1.5mu\abs{\log\sigma})$, and
$L/(\eps\abs{\log\sigma})$ is replaced by
$\sigma^{2/3}L/(\eps\sqrt{a_0}\mskip1.5mu\abs{\log\sigma})$. 

Note that in all cases, the distribution of $\lambda^0$ decays faster to the
right than to the left of $\hat\lambda$, and it is unlikely to observe
$\lambda^0$ larger than $\lc$, except when approaching the lower boundary of
Region~III. 
\end{itemiz}

In some physical applications, for instance in ferromagnets, the area
enclosed by hysteresis cycles represents the energy dissipation per period.
The distribution of the random hysteresis area can also be described, and
bounds on its expectation and variance can be obtained. We refer to
\cite{BG3} and~\cite[Section~4]{BG4} for details. 

For Stommel's box model, the above properties have two important
consequences. First, noise can drive the system from one stable equilibrium
to the other {\em before\/} the potential barrier between them disappears, so
that a smaller deviation from the mean freshwater flux than expected from
the deterministic analysis can switch the system's state. Second, this
early switching to the other state is likely only if the noise intensity
exceeds a threshold value (which is lowest when the amplitude $K$ is close 
to $\lc$). Still, the system spends roughly half of the time per period in
metastable equilibrium in the shallower well. 


\section{Delay}
\label{sec_d}

Convective motions in the atmosphere can be simulated in a laboratory
experiment known as Rayleigh--B\'enard convection. A fluid contained between
two horizontal plates is heated from below. For low heating, the fluid
remains at rest. Above a threshold, stationary convection rolls develop.
With increasing energy supply, the angular velocity of the rolls becomes
time-dependent, first periodically, and then, after a sequence of
bifurcations depending on the geometry of the set-up, chaotic. For still
stronger heating, the convection rolls are destroyed and the dynamics becomes
turbulent. 

Lorenz' famous model~\cite{Lorenz} uses a three-modes Galerkin approximation
of the hydro\-dynamic equations. The amplitudes of these modes obey the ODEs
\begin{equation}
\label{d1}
\begin{split}
\dot X &= \math{Pr} (Y-X) \\
\dot Y &= rX - Y - XZ \\
\dot Z &= -bZ + XY.
\end{split}
\end{equation}
Here $X$ measures the angular velocity of convection rolls, while $Y$ and
$Z$ parametrize the temperature field. The Prandtl number $\math{Pr}>0$ is a
characteristic of the fluid, $b$ depends on the geometry of the container,
and $r$ is proportional to the heating. 

For $0\leqs r\leqs 1$, the origin $(X,Y,Z)=(0,0,0)$ is a global attractor of
the system, corresponding to the fluid at rest. At $r=1$, this state becomes
unstable in a pitchfork bifurcation. Two new stable equilibrium branches
$(\pm\sqrt{b(r-1)},\pm\sqrt{b(r-1)},r-1)$ are created, which correspond to
convection rolls with the two possible directions of rotation. We will focus
on this simplest bifurcation, ignoring all the other sequences of
bifurcations ultimately leading to a strange attractor (see for instance
\cite{Sparrow}). 

\begin{figure}
 \centerline{\psfig{figure=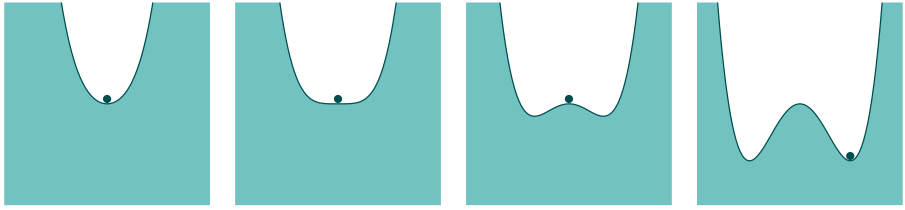,width=120mm,clip=t}}
 \captionspace
 \caption[]
 {The potential $V(x,t)=\frac14x^4 - \tfrac12 \mu(t)x^2$ transforms,
 as $\mu$ changes from negative to positive, from a single-well to a
 double-well potential.  In the deterministic case, an overdamped
 particle stays close to the saddle  for a macroscopic time before falling into one of the wells. Noise tends to reduce this delay.}
\label{fig07}
\end{figure}

We are interested in the situation where $r=r(\eps t)$ grows monotonously
through $r(0)=1$ with low speed $\eps$ (e.\,g.\ $r=1+\eps t$). Near the
bifurcation point, one can reduce the system to an invariant center
manifold, on which the dynamics is governed (c.\,f.~\cite{BK2}), after
scaling time by a factor $\eps$, by the one-dimensional equation
\begin{equation}
\label{d2}
\eps\dtot xt = \mu(t)x + c(t)x^3 + \Order{x^5}.
\end{equation}
Here $\mu(t)=a(t)+ \Order{\eps}$, where 
$a(t)=\frac12\bigbrak{-(\math{Pr}+1) + \sqrt{(\math{Pr}+1)^2 +
4\math{Pr}(r(t)-1)}\mskip1.5mu}$ is the largest eigenvalue
of the linearization of~\eqref{d1} at $0$, which has the same sign as
$r(t)-1$, and $c(t)$ is negative and bounded away from zero. The right-hand
side of~\eqref{d2} derives from a potential similar to the Ginzburg--Landau
potential~\eqref{in8} with $\lambda=0$, which remains symmetric while
transforming from a single-well to a double-well as $\mu(t)$ becomes
positive, see \figref{fig07}. 

The solution of~\eqref{d2} with initial condition $x_0>0$ for $t_0<0$ can be
written in the form 
\begin{equation}
\label{d3}
x_t = \varphi(x_0,t) \e^{\alpha(t,t_0)/\eps}, 
\qquad\qquad
\alpha(t,t_0) = \int_{t_0}^t \mu(s)\6s,
\end{equation}
with $0<\varphi(x_0,t)\leqs x_0$ for all $t$. Thus $x_t$ is exponentially
small if $\alpha(t,t_0)$ is negative. The important point to note is that
$\alpha(t,t_0)$ can be negative even when $a(t)$ is positive. For instance,
if $\mu(s)=s$, then $\alpha(t,t_0)=\frac12(t^2-t_0^2)$ is negative for
$t_0<t<-t_0$. Thus $x_t$ will remain exponentially close to the saddle at
$x=0$ up to time $-t_0$ after crossing the bifurcation point. This
phenomenon is called \defwd{bifurcation delay}. It means that when $r$ is
slowly increased, convection rolls will not appear at $r=1$, as expected
from the static analysis, but only for some larger value of $r$, which
depends on the initial condition. 

It is clear that the existence of a delay depends crucially on the fact that
$x_t$ can approach the saddle exponentially closely, where the repulsion is
very small. Noise present in the system will help kicking $x_t$ away from
the saddle, and thus reduce the delay. The question is to determine how the
delay depends on the noise intensity $\sigma$. 

For brevity, we will illustrate the results in the particular case of a
Ginzburg--Landau potential, with dynamics governed by the SDE
\begin{equation}
\label{d4}
\6x_t = \frac1\eps \bigbrak{\mu(t)x_t - x_t^3}\6t +
\frac\sigma{\sqrt\eps}\6W_t.
\end{equation}
The case without the term $-x_t^3$ has been analysed by several authors
\cite{TM,SMC,SHA,JL}, with the result that the typical bifurcation delay in
the presence of noise behaves like $\sqrt{\abs{\log\sigma}}$. The results in
\cite{BG1} cover more general nonlinearities than $-x^3$. 

We assume that $\mu(t)$ is increasing, and satisfies $\mu(0)=0$,
$\mu'(0) \geqs \text{{\it const }}>0$. For simplicity, we consider first the case where $x_t$ starts at a
time $t_0<0$ at the origin $x=0$. From the results of Section~\ref{sec_st},
we expect the paths to remain concentrated, for some time, in a set whose
width is related to the linearization of~\eqref{d4} around $x=0$. We define
the function 
\begin{equation}
\label{d5}
\bar v(t) = \bar v_0 \e^{2\alpha(t)/\eps} + \frac{\sigma^2}\eps
\int_{t_0}^t\e^{2\alpha(t,s)/\eps}\6s, 
\qquad 
\text{where $\alpha(t)=\alpha(t,0)$.}
\end{equation}
For a suitably chosen $\bar v_0\asymp\sigma^2/\abs{\mu(t_0)}$, one can
show that $\bar v(t)$ is increasing and satisfies
\begin{equation}
\label{d6}
\bar v(t) \asymp
\begin{cases}
{\sigma^2}/{\abs{\mu(t)}} 
&\text{for $t_0\leqs t\leqs-\sqrt\eps$} \\
{\sigma^2}/{\sqrt\eps} 
&\text{for $-\sqrt\eps\leqs t\leqs\sqrt\eps$} \\
{\sigma^2} \e^{2\alpha(t)/\eps} /{\sqrt\eps}
&\text{for $t\geqs\sqrt\eps$.} 
\end{cases}
\end{equation}
Note that although the curvature $\abs{\mu(t)}$ of the potential at the
origin vanishes at time $0$, $\bar v(t)$ grows slowly until time
$\sqrt\eps$ {\em after\/} the bifurcation point, and only then it starts
growing faster and faster. 

We now introduce, as in Section~\ref{sec_st}, the set 
\begin{equation}
\label{d7}
\cB(h) = \bigsetsuch{(x,t)}{\abs{x}\leqs h\sqrt{\bar v(t)}\mskip1.5mu}. 
\end{equation}
Then one can show (see~\cite[Theorem~2.10]{BG1}) the existence of a constant
$h_0>0$ such that the first-exit time $\tau_{\cB(h)}$ of $x_t$ from $\cB(h)$
satisfies 
\begin{equation}
\label{d8}
\bigprob{\tau_{\cB(h)}<t} \leqs C(t,\eps) \e^{-\kappa h^2}
\end{equation}
for all $h\leqs h_0\sigma/\bar v(t)$, where 
\begin{equation}
\label{d9}
C(t,\eps) = \frac1{\eps^2}\int_{t_0}^t \abs{\mu(s)}\6s +
\biggOrder{\frac1\eps}, 
\qquad
\text{and}
\qquad
\kappa = \frac12 - \Order{\sqrt\eps\mskip1.5mu} - \biggOrder{\frac{h^2\bar
v(t)^2}{\sigma^2}}. 
\end{equation}
The paths are concentrated in $\cB(h)$, provided $h_0^2\sigma^2/\bar v(t)^2
\geqs h^2 \gg \log C(t,\eps)$. As a consequence, we can distinguish between
three regimes, depending on noise intensity:
\begin{itemiz}
\item	{\bf Regime~I:} $\sigma\leqs\e^{-K/\eps}$ for some $K>0$.

The paths are concentrated near $x=0$ at least as long as $2\alpha(t)\ll K$.
This implies that there is still a macroscopic bifurcation delay.

\item	{\bf Regime~II:} $\e^{-1/\eps^p}\leqs\sigma\ll\sqrt\eps$ for some
$p<1$. 

The paths are concentrated near $x=0$ at least up to time $\sqrt\eps$, with
a typical spreading growing like $\sigma/\sqrt{\abs{\mu(t)}}$ for
$t\leqs-\sqrt\eps$, and remaining of order $\sigma/\eps^{1/4}$ for
$\abs{t}\leqs\sqrt\eps$.

\item	{\bf Regime~III:} $\sigma\geqs\sqrt\eps$.

The paths are concentrated near $x=0$ at least up to time $-\sigma$, with a
typical spreading growing like $\sigma/\sqrt{\abs{\mu(t)}}$. Near $t=0$, the
potential becomes too flat to counteract the diffusion, and as $t$ grows
further, paths keep switching back and forth between the wells, before
ultimately settling for a well.
\end{itemiz}

\begin{figure}
 \centerline{\psfig{figure=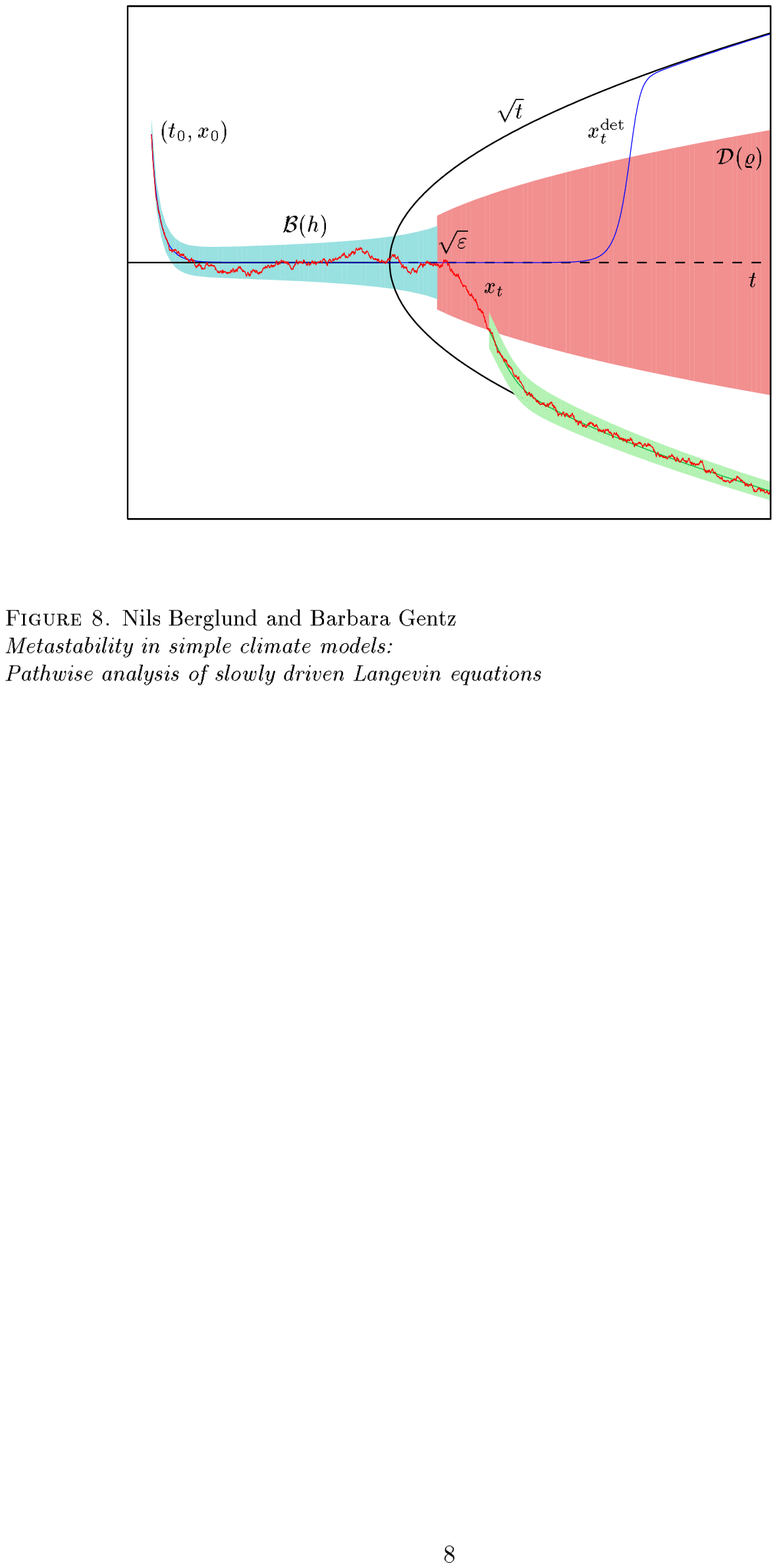,width=100mm,clip=t}}
 \captionspace
 \caption[]
 {A sample path $x_t$ of the SDE~\eqref{d4} with $\mu(t)=t$, for
 $\eps=0.01$ and $\sigma=0.015$. The deterministic solution $\xdet_t$,
 starting in $x_0>0$ at time $t_0$, jumps to the right-hand
 well, located at $x^\star(t)=\sqrt{t}$, at time $\abs{t_0}$. Typical paths
 stay in the set $\cB(h)$, whose width increases like
 $h\sigma/(\sqrt{\abs{t}}\vee\eps^{1/4})$, until time $\sqrt\eps$ after the
 bifurcation. They leave the domain $\cD(\varrho)$ (shown for
 $\varrho=2/3$) at a random time $\tau=\tau_{\cD(\varrho)}$, which is
 typically of order $\smash{\sqrt{\eps{\abs{\log\sigma}}}}$. After leaving
 $\cD(\varrho)$, each path is likely to stay in a strip of width of order
 $h\sigma/\sqrt{t}$, centred at a deterministic solution approaching
 either $+x^\star(t)$ or $-x^\star(t)$.}
\label{fig08}
\end{figure}

Similar results hold if $x_t$ starts, at $t_0<0$, away from $x=0$, say in
$x_0>0$. Then the set $\cB(h)$ is centred at the deterministic solution
$\xdet_t$ (with the same initial condition), which jumps to the right-hand
well when $\alpha(t,t_0)$ becomes positive, see~\figref{fig08}. 
In Regime~I, with $K$ sufficiently large, the majority of paths follow
$\xdet_t$ into the right-hand potential well.

It remains to understand the behaviour after time $\sqrt\eps$ in Regime~II.
To this end, we introduce the set
\begin{equation}
\label{d10}
\cD(\varrho) = \bigsetsuch{(x,t)}{t\geqs\sqrt\eps,
\abs{x}\leqs\sqrt{(1-\varrho)\mu(t)}\mskip1.5mu},
\end{equation}
depending on a parameter $\varrho\in[0,2/3)$. The set $\cD(0)$ contains the
points lying between the two stable equilibrium branches
$\pm\sqrt{\mu(t)}$. One can show (see~\cite[Theorem~2.11]{BG1}) that if
$\varrho\in(0,2/3)$ and
$\sigma\abs{\log\sigma}^{3/2}=\Order{\sqrt\eps\mskip1.5mu}$, then the
first-exit time $\tau_{\cD(\varrho)}$ of $x_t$ from $\cD(\varrho)$ satisfies 
\begin{equation}
\label{d11}
\bigprob{\tau_{\cD(\varrho)}\geqs t} \leqs C(t,\eps)
\frac{\abs{\log\sigma}}\sigma
\frac{\e^{-\varrho\alpha(t,\sqrt\eps\mskip1.5mu)/\eps}}
{\sqrt{1-\e^{-2\varrho\alpha(t,\sqrt\eps\mskip1.5mu)/\eps}}},
\end{equation}
where
\begin{equation}
\label{d12}
C(t,\eps) = \text{{\it const }} \mu(t)
\biggpar{1+\frac{\alpha(t,\sqrt\eps\mskip1.5mu)}\eps}.
\end{equation}
The estimate~\eqref{d11} shows that paths are unlikely to stay in
$\cD(\varrho)$ as soon as $t$ satisfies 
$\varrho\alpha(t,\sqrt\eps\mskip1.5mu)\gg\eps\abs{\log\sigma}$. Since $\alpha$
is quadratic in $t$, most paths will have left $\cD(\varrho)$ for 
\begin{equation}
\label{d13}
t \gg \sqrt{\eps\abs{\log\sigma}}.
\end{equation}
Once $x_t$ has left $\cD(\varrho)$, one can further show that it is likely to
track a deterministic solution which approaches the bottom of one of
the potential wells. Assume for instance that $x_t$ leaves $\cD(\varrho)$
through the upper boundary, at a random time $\tau=\tau_{\cD(\varrho)}$. 
Then, for $1/2<\varrho<2/3$,~\cite[Theorem~2.12]{BG1} shows
that the deterministic solution $\xdetof{\tau}_t$, starting in the same point
at time $\tau$, approaches the bottom of the well at $\sqrt{\mu(t)}$ like
$\eps/\mu(t)^{3/2} + \sqrt{\mu(\tau)} \e^{-\eta\alpha(t,\tau)/\eps}$, where
$\eta=2-3\varrho$, and the path $x_t$ is likely to stay in a strip of width
$\sigma/\sqrt{\mu(t)}$ around $\xdetof{\tau}$. Thus after another time span
of the form~\eqref{d13}, most paths will have concentrated near the bottom
of a potential well again.

We note that different kinds of metastability play a r\^ole here. First,
paths remain concentrated for some time near the {\em unstable\/} saddle.
Second, they will concentrate again near one of the potential wells after
some time. Some paths will choose the left-hand well and others the
right-hand well (with probability exponentially close to $1/2$ in
Regime~II), but all the paths which choose a given potential well are
unlikely to cross the barrier again. In fact, one can show that if $\mu(t)$
grows at least linearly, then the probability {\em ever\/} to cross the
saddle again is of order $\e^{-\text{{\it const}}/\sigma^2}$. If we start
the system at a {\em positive\/} $t_0$ in one of the wells, the distribution
will never approach a symmetric bimodal one.  

In the case of the Rayleigh--B\'enard convection with slowly growing heat
supply $r(\eps t)$ and additive noise, these results mean that
exponentially weak noise will not prevent the delayed appearance of
convection rolls. For moderate noise intensity, rolls will appear after a
delay of order $\sqrt{\abs{\log\sigma}/\eps}$, which is considerably
shorter than the delay in the deterministic case which is of order $1/\eps$. 
The direction of rotation is unlikely to change after another time
span of that order. For strong noise, convection rolls may appear
early, but their angular velocity will fluctuate around zero until a
time of order $\sigma/\eps$ after the bifurcation before settling for a
sign, and even then occasional changes of rotation direction are possible. 


\subsection*{Acknowledgements}
We thank the organisers for the invitation to Chorin and the
opportunity to present our results during the {\em Second Workshop on
Stochastic Climate Models}. We enjoyed stimulating discussions in a
pleasant atmosphere.
  
\goodbreak


\small
\bibliography{../BG}
\bibliographystyle{abbrv}		

\goodbreak

\bigskip\bigskip\noindent
{\small 
Nils Berglund \\ 
{\sc Department of Mathematics, ETH Z\"urich} \\ 
ETH Zentrum, 8092~Z\"urich, Switzerland \\
{\it E-mail address:\/ }{\tt berglund@math.ethz.ch}

\bigskip\noindent
Barbara Gentz \\ 
{\sc Weierstra\ss\ Institute for Applied Analysis and Stochastics} \\
Mohrenstra{\ss}e~39, 10117~Berlin, Germany \\
{\it E-mail address:\/ }{\tt gentz@wias-berlin.de}
}


\end{document}